\documentclass[onecolumn,preprintnumbers,elsart]{revtex4}
\usepackage{makeidx}
\usepackage{amssymb}
\usepackage{amsmath}
\usepackage{mathrsfs}
\usepackage{graphicx}
\usepackage{dcolumn}
\usepackage{bm}
\usepackage[center]{subfigure}
\usepackage{color}

\begin{document}

\title{Anderson localization in metallic nanoparticle arrays }
\author{Zhijie Mai,$^{1}$ Fang Lin,$^{1}$ Wei Pang,$^{2}$ Haitao Xu,$^{1}$ Suiyan Tan,$^{1}$ Shenhe Fu,$^{3,\ast}$} 
\email{shenhe_fu@163.com}
\author{Yongyao Li$^{1}$}
\affiliation{$^{1}$Department of Applied Physics, South China Agricultural University, Guangzhou 510642, China\\
$^{2}$Department of Experiment Teaching, Guangdong University of Technology, Guangzhou 510006, China\\
$^{3}$State Key Laboratory of Optoelectronic Materials and Technologies,\\
School of Physics and Engineering, Sun Yat-sen University, Guangzhou 510275, China}

\begin{abstract}
Anderson localization has been observed in various types of waves,
such as matter waves, optical waves and acoustic waves. Here we reveal that the effect of Anderson localization can be also induced in metallic nonlinear nanoparticle arrays excited by a random electrically driving field. We find that the dipole-induced nonlinearity results in ballistic expansion of dipole intensity during evolution; while the randomness of the external driving field can suppress such an expansion. Increasing the strength of randomness above the threshold value, a localized pattern of dipole intensity can be generated in the metallic nanoparticle arrays. By means of statistics, the mean intensity distribution of the dipoles reveals the formation of Anderson localization. We further show that the generated Anderson localization is highly confined, with its size down to the scale of incident wavelength. The reported results might facilitate the manipulations of electromagnetic fields in the scale of wavelength.\\
\textbf{OCIS:}(190.1450) Bistability; (190.6135) Spatial solitons; (250.5403) Plasmonics.
\end{abstract}
\maketitle

\section{Introduction}

Anderson localization (AL) is a fundamental wave phenomenon that occurs in
strongly disordered media \cite{anderson}. In 1958, P. W. Anderson
first suggested the possibility of electron localization inside a
semiconductor, where the essential randomness is introduced \cite{anderson}. To
date, it has been reported that AL has been observed in various types of waves, such as matter waves\cite{billy, roati, ludlam, chabe}, optical waves \cite{scheffold, schwartz, karbasi2012, karbasi2013}, and acoustic waves \cite{hu}.

Due to the diffraction limit of optical waves, the realization of
optical AL at the extremely narrow scale is generally
difficult to achieve in dielectric media, such as photonic lattices
and optical fibres \cite{scheffold, schwartz,
  karbasi2012, karbasi2013}. However, it has been suggested that
surface plasmon polaritons, which travel along a metal-dielectric
interface, are a highly confined electromagnetic field with a size smaller than the wavelength. Therefore, to realize highly confined AL, an important candidate is based on plasmonics. Recently, the coupling
plasmonic waveguide system\cite{shi, deng} and the graphene plasmonic
waveguide system\cite{xu} were proposed for producing the extremely narrow AL by introducing random potential into these waveguide systems.

It has been reported that the AL of electromagnetic polar waves in a random linear chain of dipoles has been studied \cite{PRB83}. In this work, we propose the realization of AL of surface plasmon polaritons in the arrays of
metallic nanoparticles. Using a homogeneous electrically driving field, the nonlinear modulation instability and the bistability of
optically induced dipoles in the metallic nanoparticle arrays have been
studied in detail \cite{prl, oe, sr}. Here, to realize AL of surface plasmon polaritons, a random electrically
driving field, which has been used to create optical lattices with random perturbations \cite{schwartz}, is utilized to excited the metallic nanoparticle
arrays. Our numerical simulations demonstrate that the generated AL is highly confined, with its size down to the scale of incident wavelength.

The remainder of this paper is organized as follows. In section II, we
introduce the nonlinearly coupling equations that describe the dynamics
of particle's dipoles excited by the external field in the arrays of metallic nanoparticles. In section
III, we study the dynamics of the bright plasmonic dipole mode induced by the homogeneous driving field. In section IV, we investigate the generation of AL of dipole intensity by introducing the random electrically driving field. In section V, we investigate the size of the generated AL in the present system, finding that the size of the AL can be reduced to the scale of a wavelength of the excited field. The paper is concluded in section VI. The appendix presents the detailed parameters of the model.

\section{Model and basic equations}
In the nanoparticle arrays, each spherical silver nanoparticle is arrayed linearly equidistant and embedded in a SiO$_{2}$ host, as shown in Fig. 1. The appendix presents the specific parameters of these silver nanoparticles. Surface plasmon polaritons could be excited by the electrical field imposed on these nanoparticle arrays, and the dynamics of such excited surface plasmon polaritons can be modelled using the following coupling equations \cite{prl,oe,sr}
\begin{figure}[t]
\centering\includegraphics[width=10cm,height=5cm]{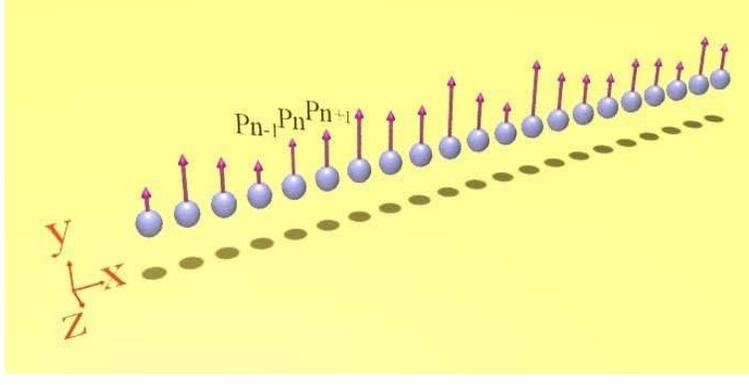}\label{fig1}
\caption{Schematic illustration of an array of
  spherical silver nanoparticles embedded in a SiO$_{2}$ host; the
  radius of the sphere is $a$, and the center-to-center distance of two adjacent particles is $d$. The dipoles are induced by an electrical driving field.}
\end{figure}

\begin{eqnarray}
-i\frac{dP_{n}^{\perp}}{d\tau}+(-i\gamma+\Omega+|P_{n}|^{2})P_{n}^{\perp}+\sum_{m\neq n}G_{n,m}^{\perp}P_{m}^{\perp}&=&E_{n}^{\perp}, \label{diff1} \\
-i\frac{dP_{n}^{\parallel}}{d\tau}+(-i\gamma+\Omega+|P_{n}|^{2})P_{n}^{\parallel}+\sum_{m\neq n}G_{n,m}^{\parallel}P_{m}^{\parallel}&=&E_{n}^{\parallel}, \label{diff2}
\end{eqnarray}
where $P_{n}^{\bot,
  \parallel}=p_{n}^{\perp,\parallel}\sqrt{\chi^{(3)}}/(\sqrt{2(\varepsilon_{\infty}+2\varepsilon_{h})}\varepsilon_{h}a^3)$
are the dimensionless slowly varying amplitudes of the vertical and
parallel dipoles of the $n^{\mathrm{th}}$ particle, respectively. The
indices `$\perp$' and `$\parallel$' represent the vertical and
parallel directions with respect to the array axis,
respectively. Thus, the total intensity of the dipole for the
$n^{\mathrm{th}}$ particle is given as
$|P_{n}|^2=|P_{n}^{\perp}|^2+|P_{n}^{\parallel}|^2$. $\gamma=\nu/(2\omega_{0})+(k_{0}a)^3\varepsilon_{h}/(\varepsilon_{\infty}+2\varepsilon_{h})$
is the scaled damping, with
$k_{0}=\omega_{0}/c\sqrt{\varepsilon_{h}}$. $\Omega=(\omega-\omega_{0})/\omega_{0}$
is the detuning frequency of the dipoles. For a wavelength of $400nm$,
$\Omega=0$. $\tau=\omega_{0}t$ is the scaled elapsed
time. $E_{n}^{\perp,
  \parallel}=-3\varepsilon_{h}\sqrt{\chi^{(3)}}E_{n}^{ex, \perp,
  \parallel}/\sqrt{8(\varepsilon_{\infty}+2\varepsilon_{h})^3}$
represent the slowly varying amplitudes of the external optical fields
in the respective directions. $G_{n,m}^{\perp,\parallel}$ is the
linearly coupling parameter between the $n^{\mathrm{th}}$ and
$m^{\mathrm{th}}$ particles in the corresponding directions and is
induced by the long-range dipole-dipole interactions. According to
\cite{prl}, $G_{n,m}^{\perp,\parallel}$ can be expressed as follows:
\begin{eqnarray}
G_{n,m}^{\perp}&=&\frac{\eta}{2}\left[(k_{0}d)^{2}-\frac{ik_{0}d}{|n-m|}-\frac{1}{|n-m|^2}\right]\frac{e^{-ik_{0}d|n-m|}}{|n-m|},\\
G_{n,m}^{\parallel}&=&\eta\left(\frac{ik_{0}d}{|n-m|}+\frac{1}{|n-m|^2}\right)\frac{e^{-ik_{0}d|n-m|}}{|n-m|},
\end{eqnarray}
where $\eta=\frac{3\varepsilon_{h}}{\varepsilon_{\infty}+2\varepsilon_{h}}(\frac{a}{d})^3$. Note that Eqs. (\ref{diff1}) and (\ref{diff2}) are suitable for cases of finite and infinite nanoparticle chains. The real-time evolution methods can be used to solve the $P_{n}$ in Eqs. (\ref{diff1}) and (\ref{diff2}).

For simplification, the optically induced dipole in the vertical
direction (i.e., the character of `$\perp$') is considered, while
neglecting the induced dipole that is polarized along the parallel
direction. Therefore, the dipole of the n$^{th}$ particle can be
written as
$\mathbf{P}_{n}=(P^{\perp}_{n},P^{\parallel}_{n})=(P_{n},0)$. It was
shown that in stationary conditions, the induced dipole exhibits a
standard bistable curve that consists of three branches, namely, the low, middle, and top branches\cite{oe}. It was suggested that the dipole evolution dynamics is stable when the dipole is initially located at the low and top branches; however, it becomes unstable when the dipole is initially located at the middle branch\cite{oe}.

\section{Dynamics of bright plasmonic dipole mode}
We previously demonstrated that \cite{mai}, using a homogeneous
electrically driving field, a dipole kink could be formed when two
adjacent particles' dipole intensities were initially located at the
low and top branches with respect to the bistable curve. Localized
dipole modes could also be formed when two kinks were created in the
nanoparticle arrays. As the intensity of the external
field ($|E_{0}|^2$) is increased beyond the threshold, the kink begins
to move, leading to the dipole intensity ($|Pv|^2$) jumping from the
low branch to the top branch, thereby breaking the stability of the dipole mode \cite{mai}.
\begin{figure}[t]
\centering
\subfigure[]{\includegraphics[width=5cm, height=4cm]{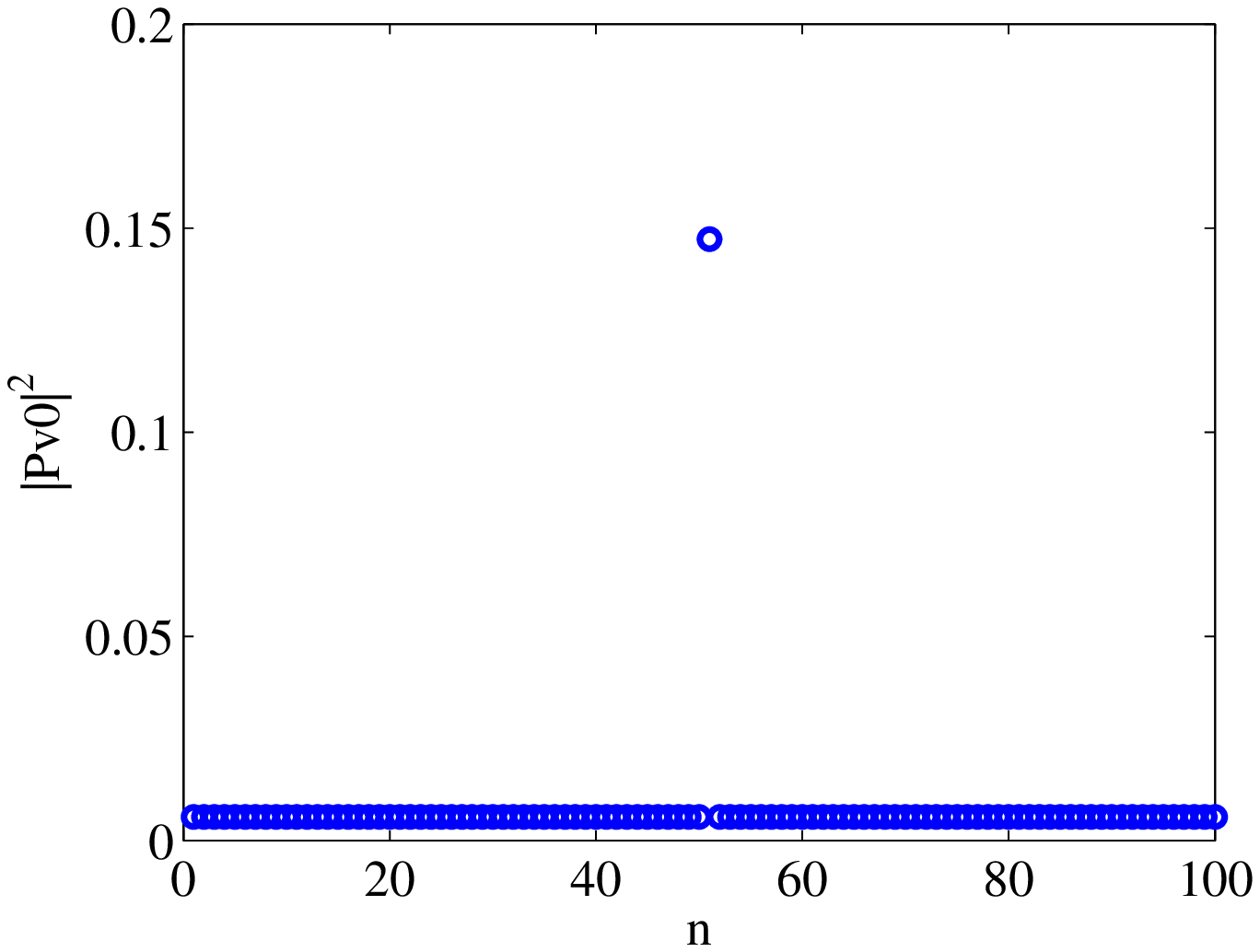}}\label{fig2a}
\subfigure[]{\includegraphics[width=5cm, height=4cm]{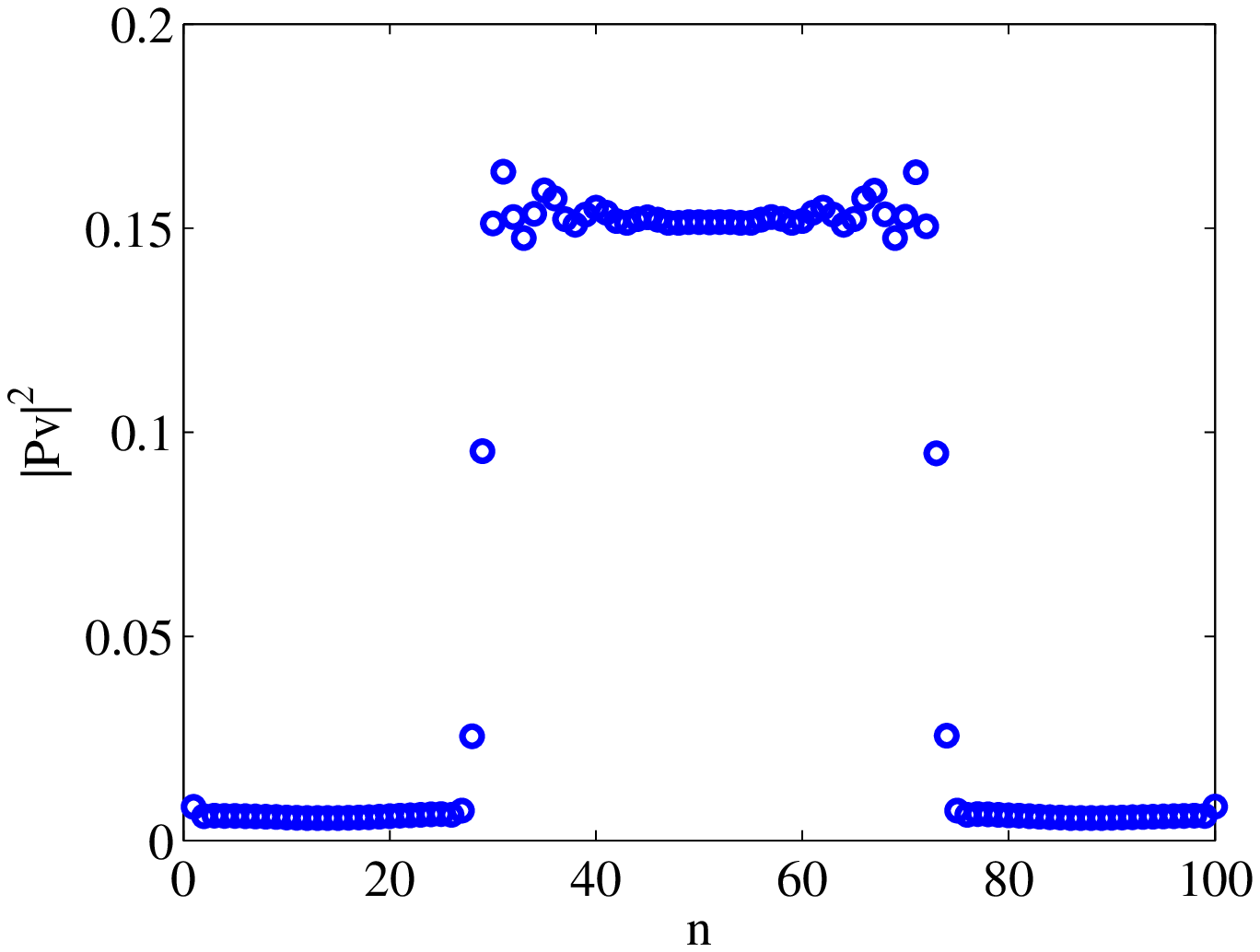}}\label{fig2b}
\subfigure[]{\includegraphics[width=5cm, height=4cm]{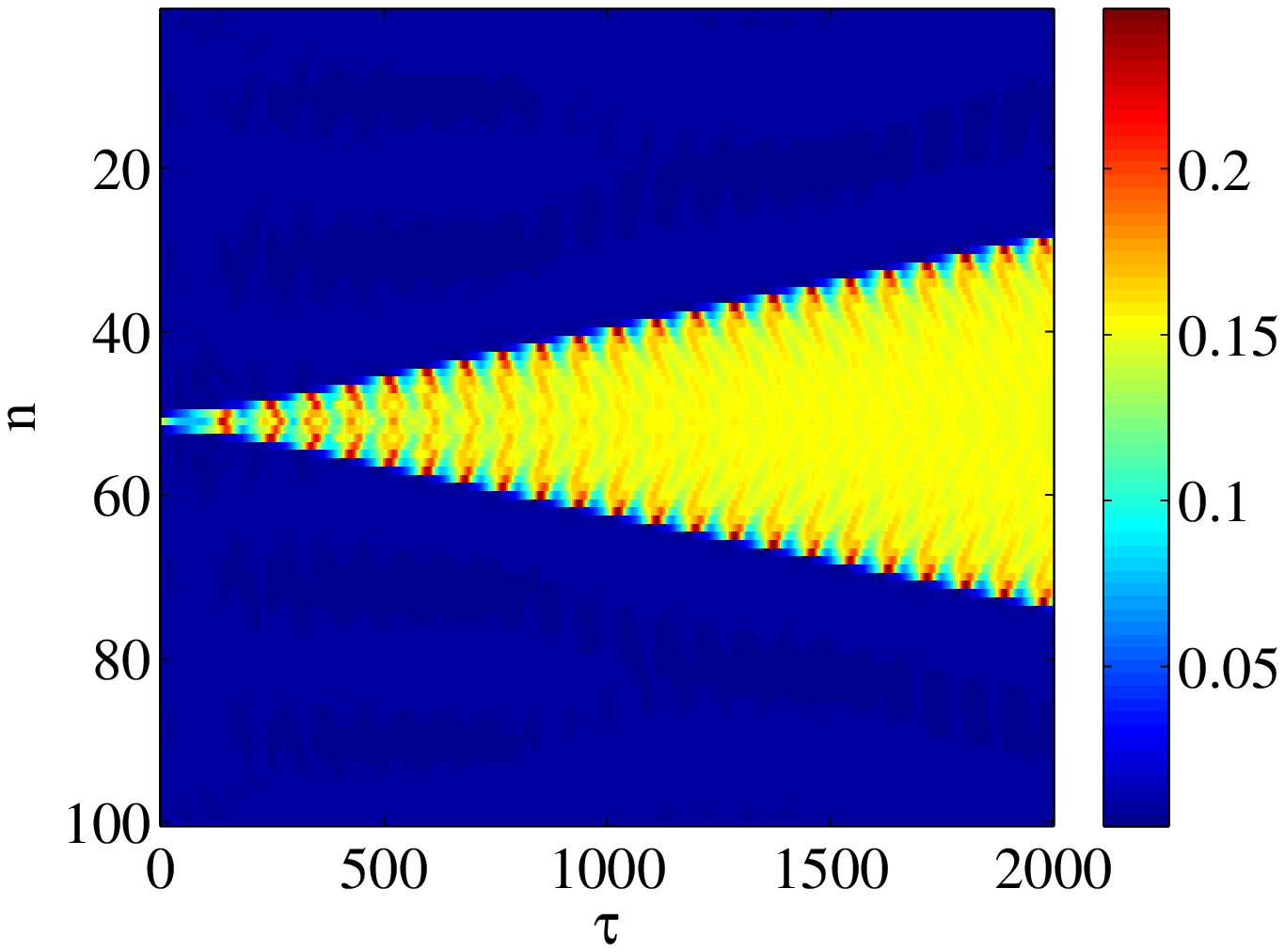}}\label{fig2c}
\subfigure[]{\includegraphics[width=5cm, height=4cm]{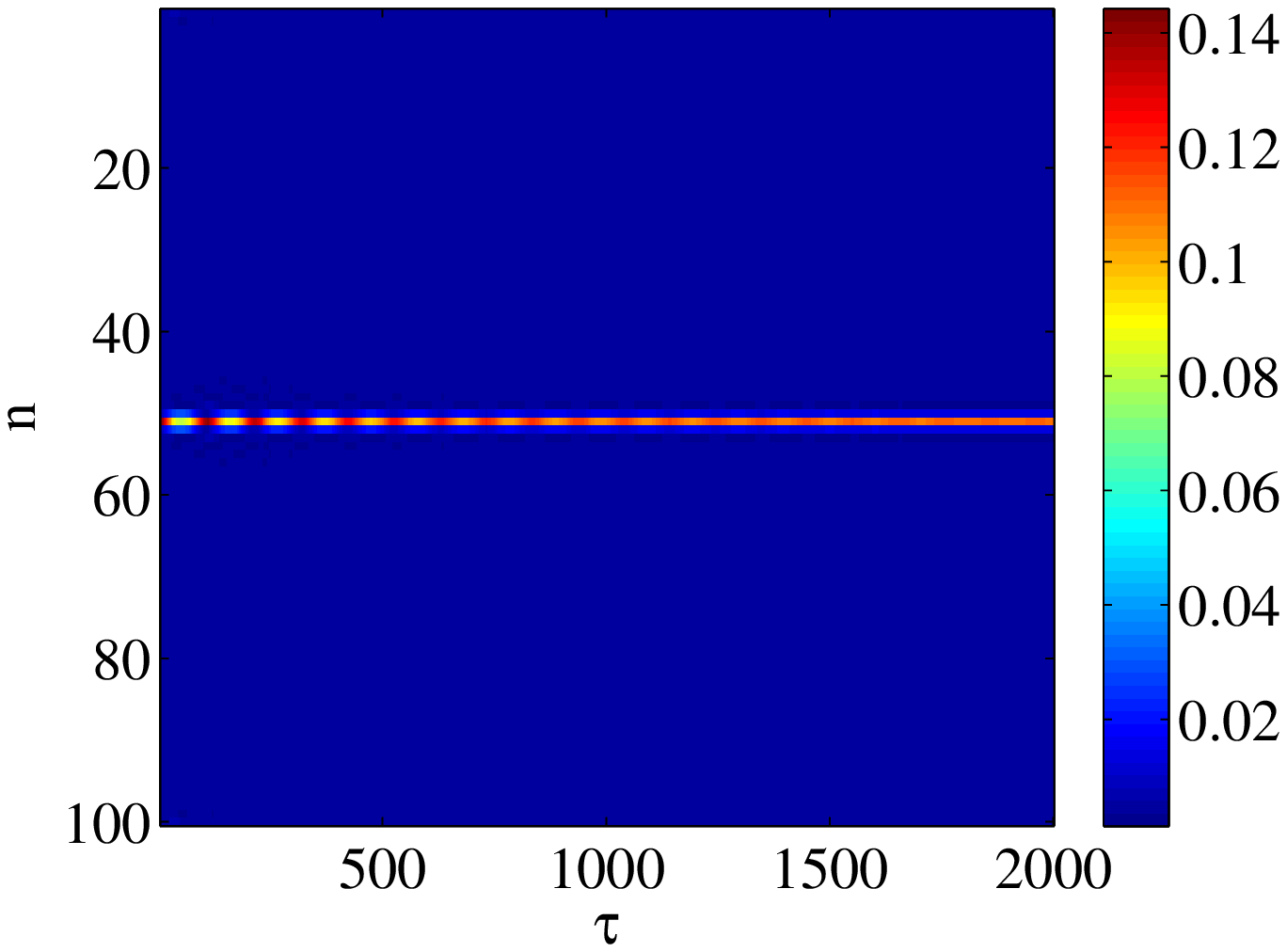}}\label{fig2d}
\caption{Bright plasmonic dipole mode in a 100-particle
  system excited by a homogeneous electric field of
  $|E_{0}|^2=0.90\times10^{-4}$ under $\Omega=-0.1$ (the incident
  wavelength is approximately 440 nm): (a) initial dipole intensity at
  $\tau=0$, (b) dipole intensity at $\tau=2000$, (c) the evolution of
  the dipole intensity under and excitation of
  $|E_{0}|^2=0.90\times10^{-4}$, and (d) the evolution of the  dipole
  intensity under and excitation of $|E_{0}|^2=0.6\times10^{-4}$,
  which is below the threshold value $|E_{c}|^2$.}
\end{figure}

Here we produce the bright dipole modes, as shown in Figs. 2(a)-2(c). As mentioned, this can
be achieved by using the homogeneous external field and initially
setting one of the particle's dipole intensity at the top branch,
while the other particle's dipole is at the low branch. In our
simulation, we set the external field as
$|E_{0}|^2=0.90\times10^{-4}$, and the direction of the driving field
is vertical to the array direction. Figure 2(a) shows the corresponding
initial state of the dipole intensities, while Fig. 2(b) illustrates
the dipole intensity after the real-time evolution of $\tau=2000$. It
clearly shows that the width of the bright dipole mode is broadening
during the dipole evolution. This result means that, during the course
of evolution, the dipoles of more particles jump from the low branch
to the top branch. Figure 2(c) shows the real-time evolution of the
dipole intensity. The blue-coloured area represents the intensity of
dipoles that are located at the low branch, while the yellow-coloured
area corresponds to the intensity of dipoles that are located at the
top branch. The broadening behaviour of the dipole mode illustrated in
Fig. 2(c) is similar to the motion of shock waves that
was observed in the discrete nonlinear dissipative medium
\cite{PRE62}. According to Ref. \cite{oe}, the speed of the kink is
related to $|E_{0}|^2$: the larger $|E_{0}|^2$ is, the faster is the
speed of the kink. Additionally, note that there is a threshold of
$|E_{0}|^2$ for the kink's motion. In the present system, the
threshold value can be achieved as $|E_{c}|^2=0.65\times10^{-4}$. Below
this threshold, the kink stops, as shown in Fig. 2(d).

\section{Anderson localization of plasmonic dipole mode}
To observe AL of surface plasmon polaritons in the
present nanoparticle arrays, we introduce a random electrical driving
field to excite the plasmonic dipoles. The external random electrical
fields consist of a series of independent electrical fields. Each
electrical field has a size of 40 nm and is successively imposed on
the nanoparticle arrays. Note that such a narrow electrical driving
field is achievable since recent work has reported
that the size of the light spot can be reduced to a scale of
less than 0.1$\lambda^2$ \cite{xiexiangsheng}. By optimizing the system, as well as the modulations of optical field, a narrower
light spot is expected \cite{xie2014}.

In the simulations, the varying range of the external driving fields
that are imposed on the 100-particle system is set as
$|E_{0}|^2=0.45\times10^{-4}\sim1.35\times10^{-4}$. It was demonstrated that the induced dipole intensity relies on the intensity of the driving field \cite{mai}. Therefore, for a random electrical driving field, the induced dipole intensity of each particle is different from that induced from other particles. Hence, the excited dipoles of the nanoparticle arrays can be considered as a random system, which provides possibility for the generation of localized dipole modes. To observe this effect, the dipole intensity of the middle nanoparticle is set to the top branch; see Fig. 3(a), while the dipole
intensities of the other nanoparticles are located at the low
branch. Figure 3(b) depicts the intensity distribution after the real-time
evolution of $\tau=2000$. Figure 3(c) shows the evolution dynamics of
the dipole intensity. It shows that the pattern still
broadens in the short stage, but it becomes localized for a longer
propagation stage, indicating the formation of a stable localized bright dipole
mode. This behaviour is very different from that observed in
Figs. 2(a)-2(c). We attribute this phenomenon of a localized pattern to the introduction of a random electrical driving
field. Owing to the randomness of the driving field, for the driving intensity $|E_{0}|^2$ below the threshold, the dipole-induced nonlinearity is negligible, and the dipole evolution becomes stable, which helps to suppress the expansion of dipole intensity.

\begin{figure}[htbp]
\centering
\subfigure[]{\includegraphics[width=5cm, height=4cm]{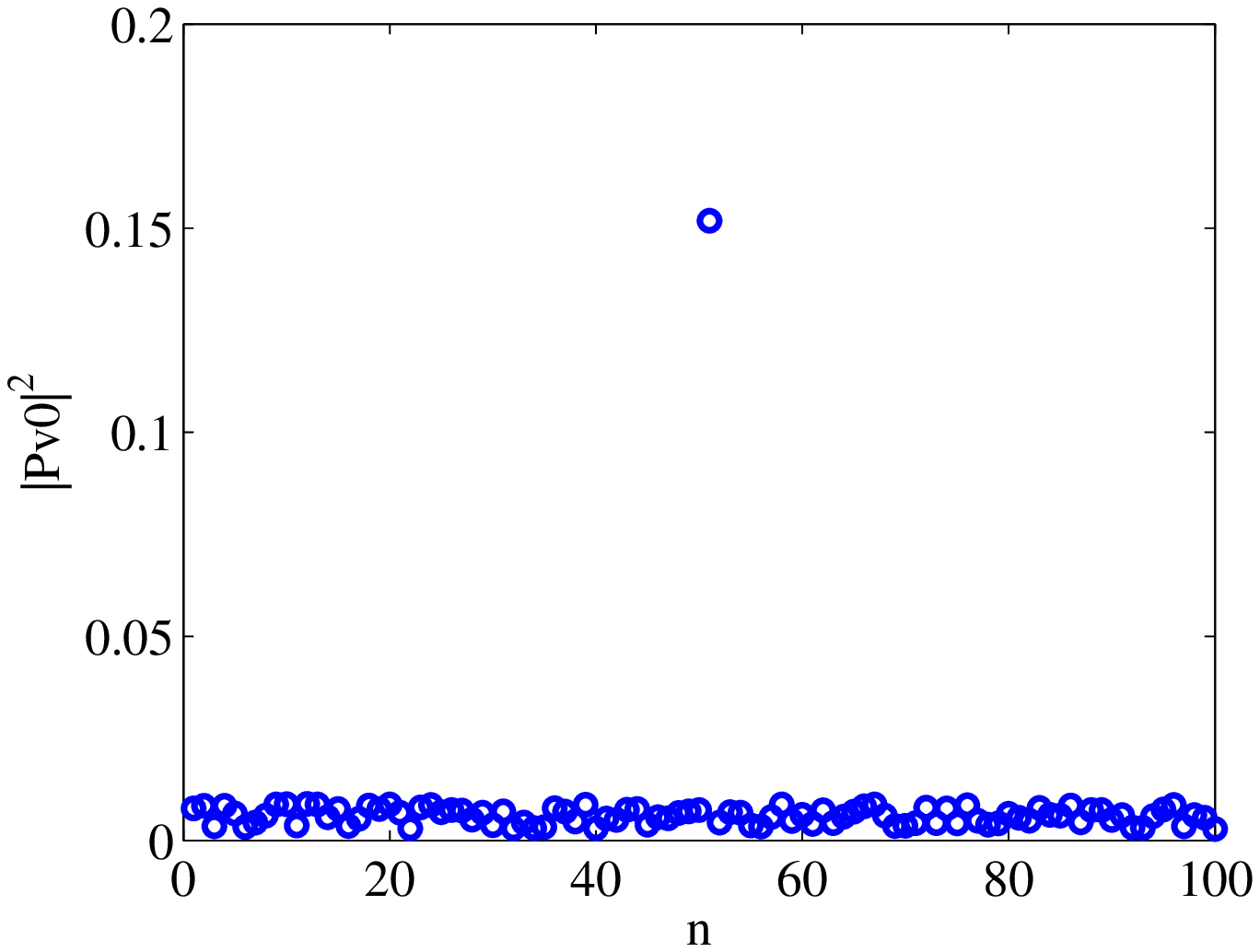}}\label{fig3a}
\subfigure[]{\includegraphics[width=5cm, height=4cm]{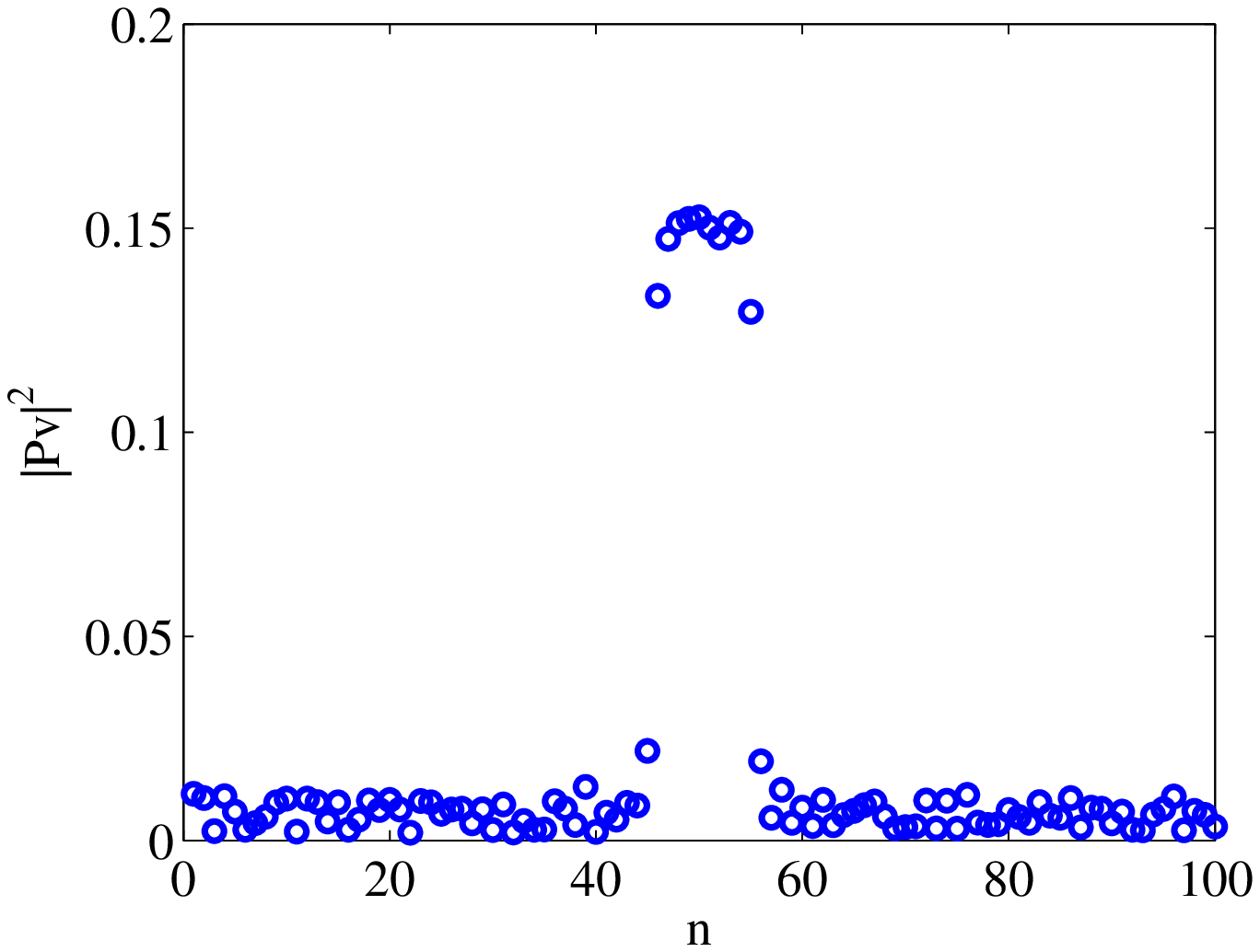}}\label{fig3b}
\subfigure[]{\includegraphics[width=5cm, height=4cm]{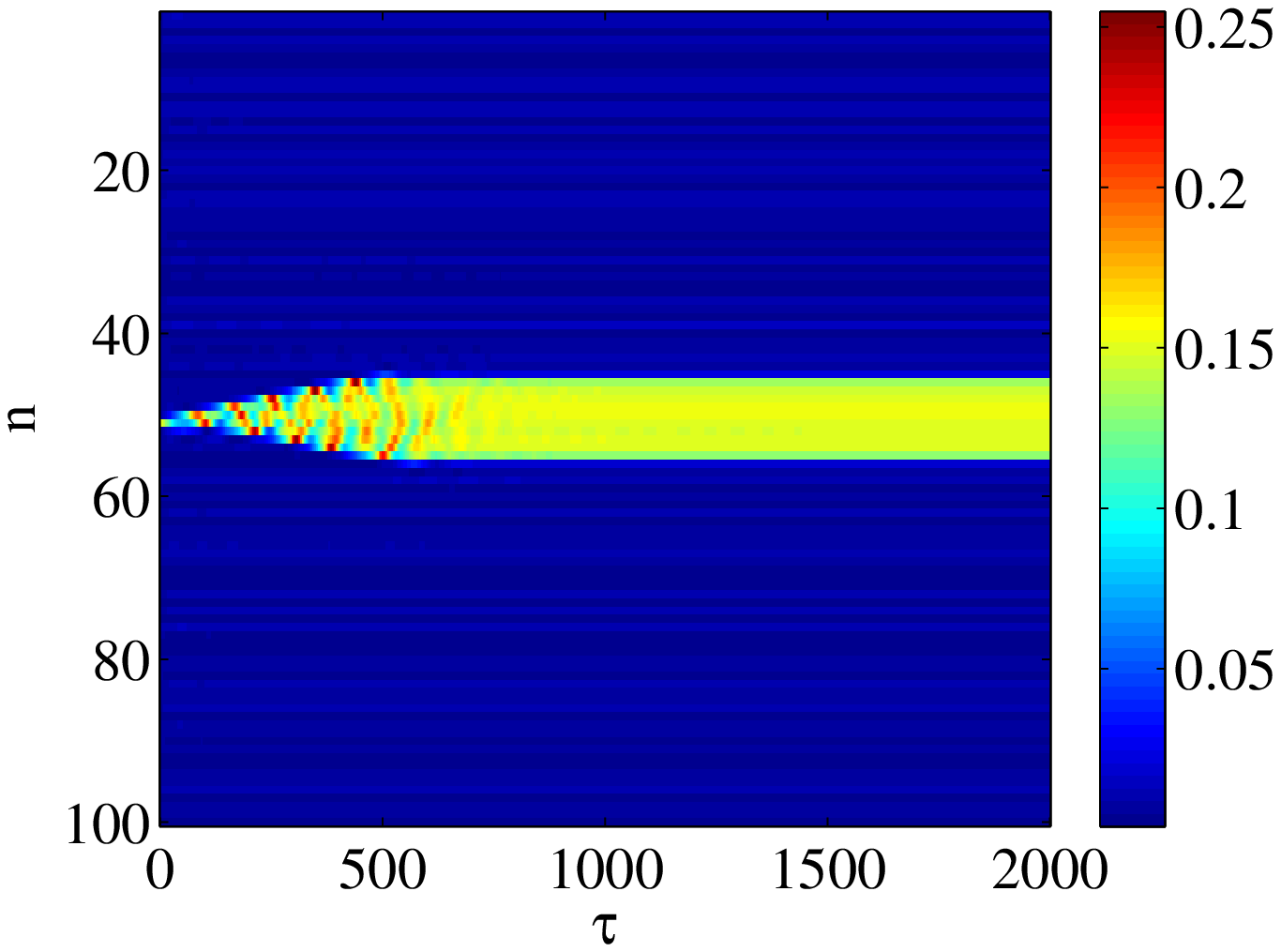}}\label{fig3c}
\caption{Localized plasmonic dipole mode in a 100-particle
  system excited by the random electrical driving field under
  $\Omega=-0.1$: (a) initial dipole intensity distribution at $\tau=0$, (b) the
  dipole intensity distribution at $\tau=2000$, and (c) the evolution of dipole intensity.}
\end{figure}

To investigate the influence of statistical regularities of the random
electrical driving field on the formation of localized dipole modes, we
perform 200 real-time evolutions of dipole modes excited by the random
electrical field. For each real-time simulation, the electrical
field imposed on the nanoparticles is different, but the varying range
of the driving field is kept the same, i.e., set as
$|E_{0}|^2=0.45\times10^{-4}\sim1.35\times10^{-4}$. In this case, we
can observe through simulations the phenomenon of AL of the surface plasmon polaritons, as evident in
Fig. 4. Figure 4(a) shows the mean value of the dipole intensity after
real-time evolution at $\tau=2000$, with the detuning parameter set as
$\Omega=-0.1$. The blue scattered circle denotes the calculated mean
dipole intensity at $\tau=2000$; while the red solid curve shows the
fitting results. Figure 4(b) illustrates the evolution of the dipole
intensity modes. Both Figs. 4(a) and 4(b) clearly depict a nice
localized dipole mode, i.e., Anderson localization.  Note that Fig. 4
is the statistical results after 200 real-time evolutions, but each
simulation, e.g., see Fig. 3, can produce a localized dipole pattern.
We should point out that the observed localized pattern shown in Fig. 4 is not caused by the dipole-induced nonlinearity. This can be verified by the results shown in Fig. 2. Due to the randomness of the driving field, for the driving intensity below the threshold value, the dipole intensity $|P_{v}|^2$ is invariant with time (it is stable during evolution). Under this circumstance, the nonlinear model described by Eqs. (1) and (2) can be reduced to a linear system. As a result, the randomness of the driving field, as well as the long-range dipole-dipole interactions, helps to suppress the dipole expansion, eventually leading to Anderson localization.

\begin{figure}[htbp]
\centering
\subfigure[]{\includegraphics[width=6cm]{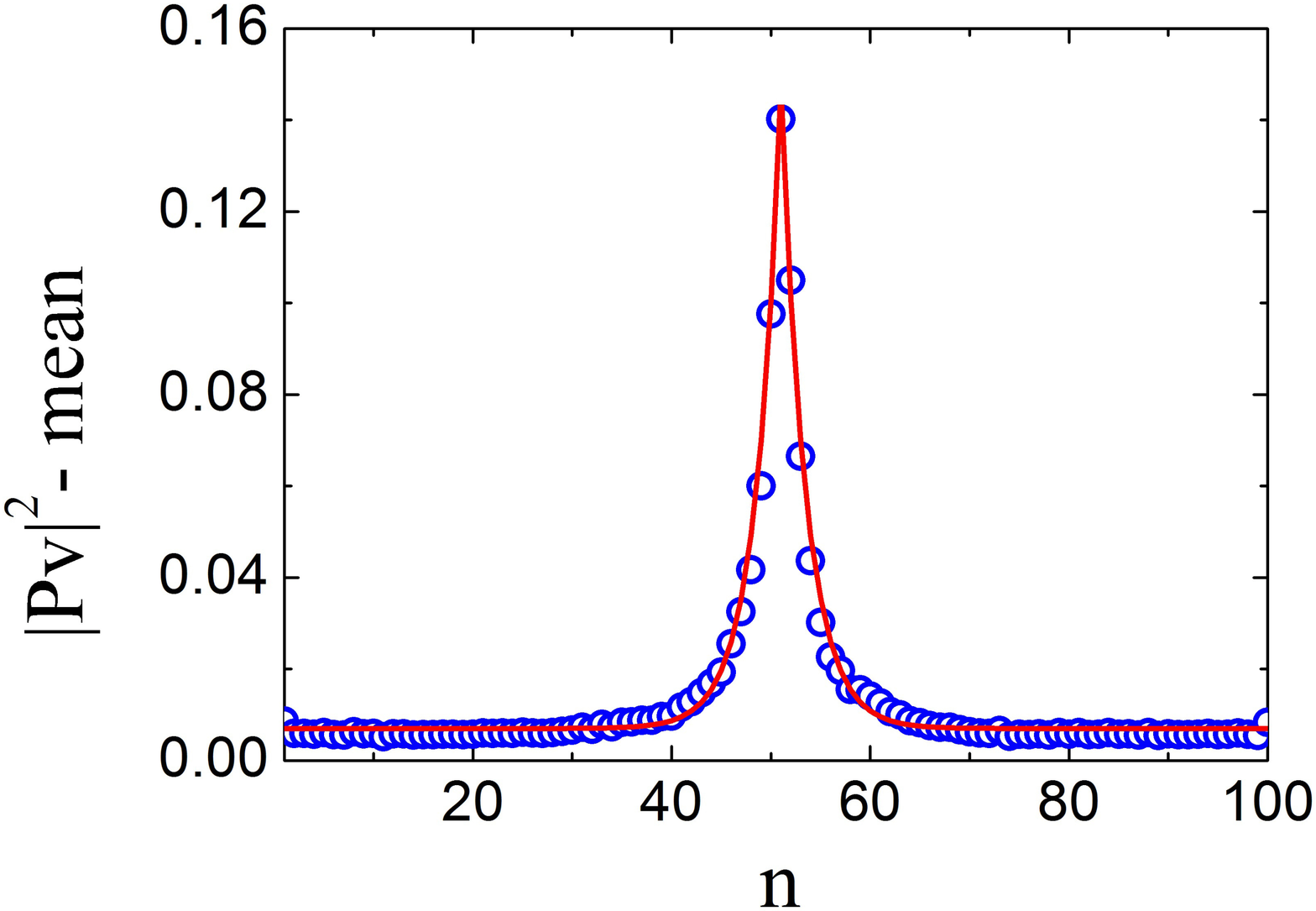}}\label{fig4a}
\subfigure[]{\includegraphics[width=6cm]{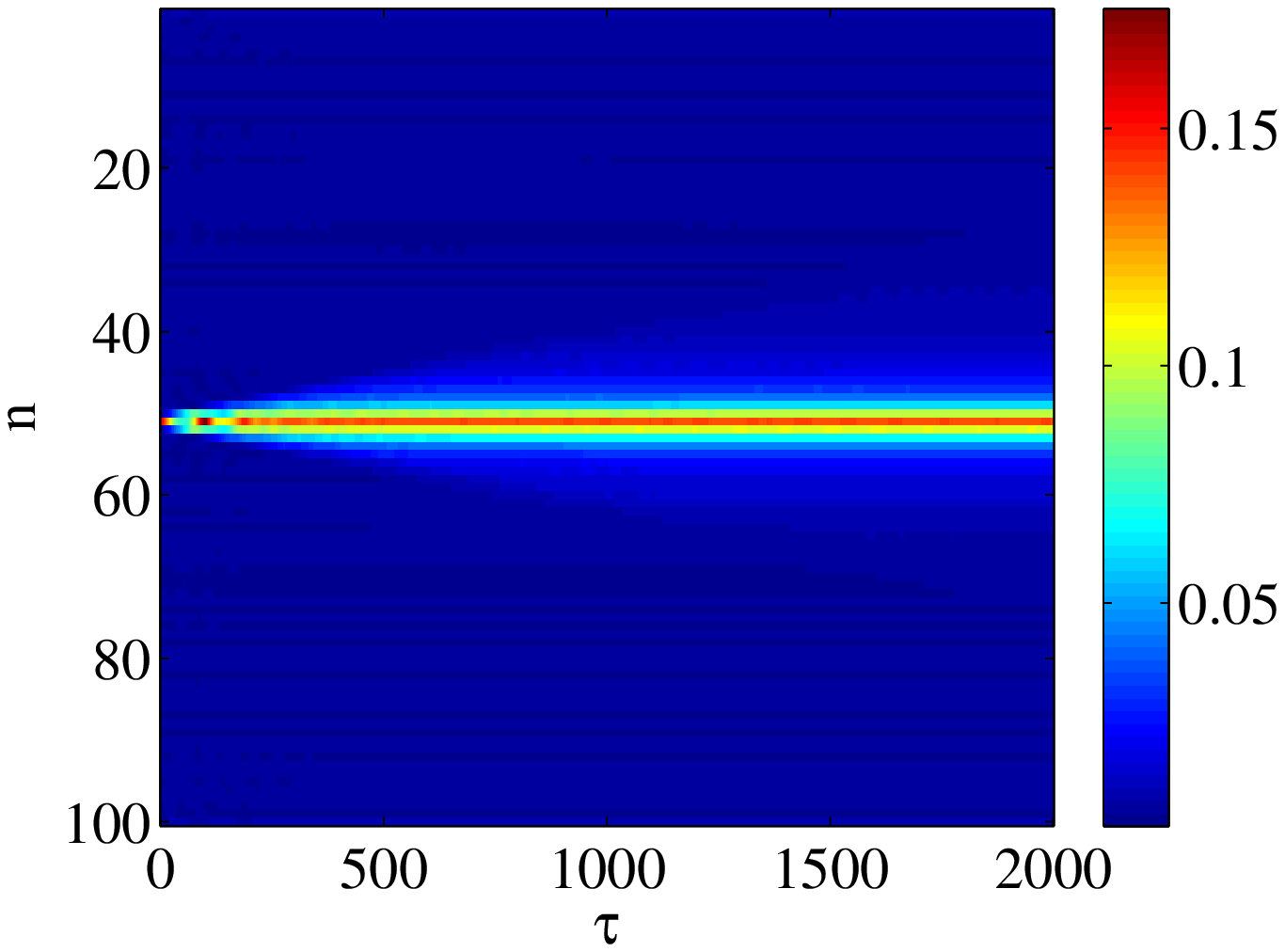}}\label{fig4b}
\caption{The statistical results of 200 real-time evolutions of
  dipole intensity excited by the random electrical driving field
  under $\Omega=-0.1$. The varying range intensity of the random driving field for each simulation is kept the same, i.e., set as
  $|E_{0}|^2=0.45\times10^{-4}\sim1.35\times10^{-4}$: (a) Mean distribution of dipole intensity at $\tau=2000$. The blue circles denote the calculated mean value;
  while the red curve represents the fitting result of $0.14exp[\pm(x-51)/2.5]+0.007$. (b) The evolution dynamics of the mean dipole intensity.}
\end{figure}

Next, we investigate the relation between the strength of randomness of
the electrical driving field and the formation of AL. For this purpose, a random factor $\kappa$ that describes the strength of the randomness is introduced, as follows:
\begin{equation}
{\left| E \right|}^{2}={\left| E_{0} \right|}^{2}\pm\kappa{\left| E_{0} \right|}^{2},
\label{kappa}
\end{equation}
Equation (5) indicates that the electrical driving field varies from ${\left|
    E_{0} \right|}^{2}-\kappa{\left| E_{0} \right|}^{2}$ to ${\left|
    E_{0} \right|}^{2}+\kappa{\left| E_{0} \right|}^{2}$. Increasing the value of $\kappa$ will broaden the varying range of the
random electrical driving field, and hence enhances the strength of the randomness. As a result, the probability that the
intensity of the driving field is below the threshold intensity
$|E_{c}|^2$ is also increased.

To observe the effect of $\kappa$ on the formation of AL, we perform simulations with different values of
$\kappa$, as shown in Fig. 5. Figures 5(a)-5(c) present the results
of a single real-time evolution with (a) $\kappa$=0.2,
(b) $\kappa$=0.4, and (c) $\kappa$=0.6, respectively. From
Figs. 5(a)-5(c), we can observe that the width of the bright
plasmonic dipole mode decreases as the strength of the randomness $\kappa$
increases. In particular, for the case of $\kappa=0.4,0.6$, see
Figs. 5(b) and 5(c), we can observe the formation of a stable localized dipole pattern. To observe the phenomenon of AL, we
perform the statistical calculation of such dipole intensity after 200
real-time evolutions, keeping the parameter $\kappa$ unchanged. The
mean values of the statistical results are shown in
Figs. 5(d)-5(f). From these numerical results, it is shown that the
effect of AL becomes stronger with a larger $\kappa$. Note that for each real-time
simulation, an evolved localized pattern induced by the random
electrical field was observed, e.g., see Fig. 3. However, by means of
statistics, the mean intensity distribution of $|Pv|^2$ after 200
real-time simulations reveals the phenomenon of AL,
as shown in Fig. 5. Furthermore, it is shown that for a larger $\kappa$, the probability that the external driving intensity is less than the threshold value becomes larger, which leads to a narrower localized  pattern.

\begin{figure}[htbp]
\centering
\subfigure[]{\includegraphics[width=4.3cm]{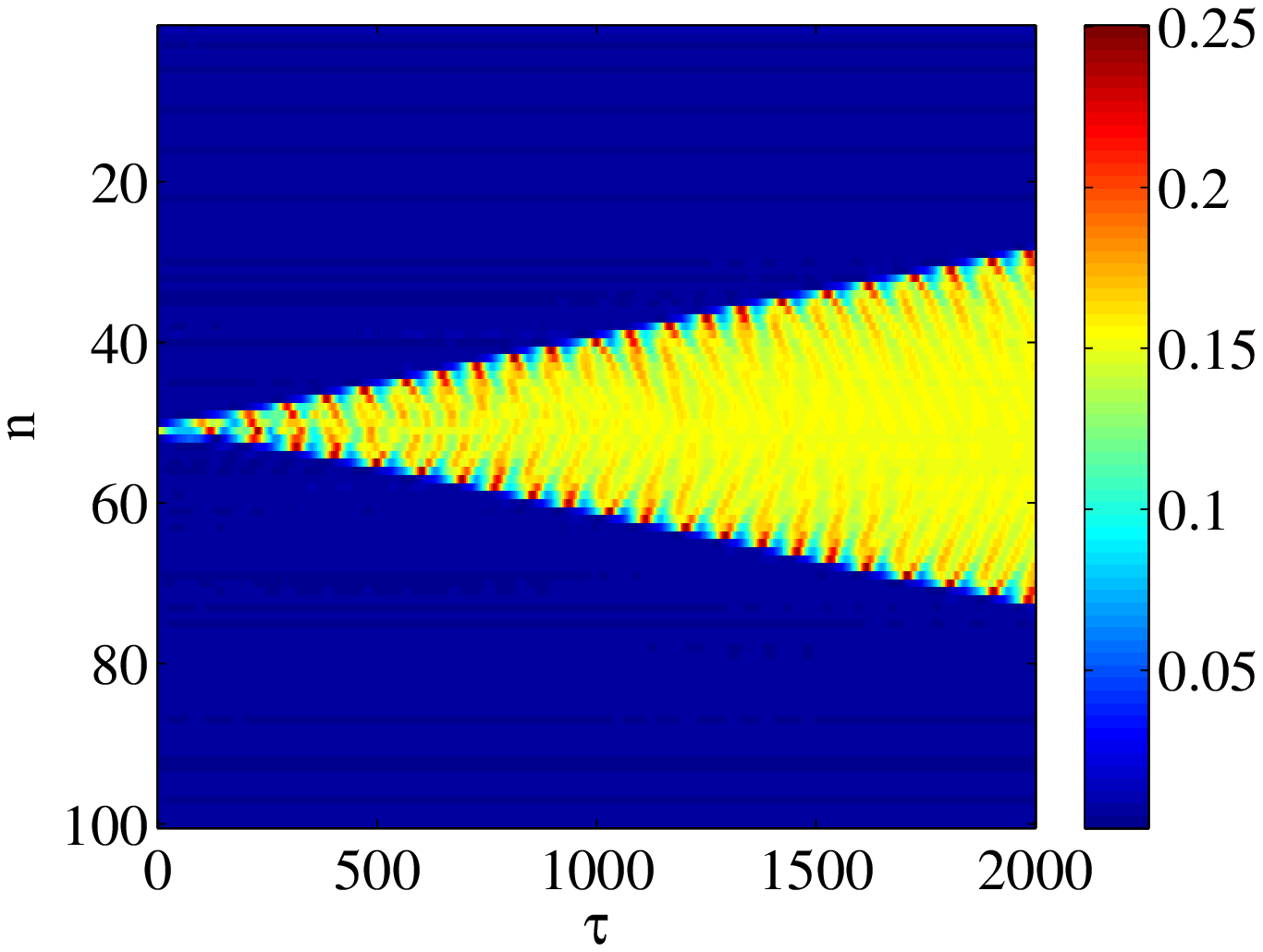}}\label{kappa2}
\subfigure[]{\includegraphics[width=4.3cm]{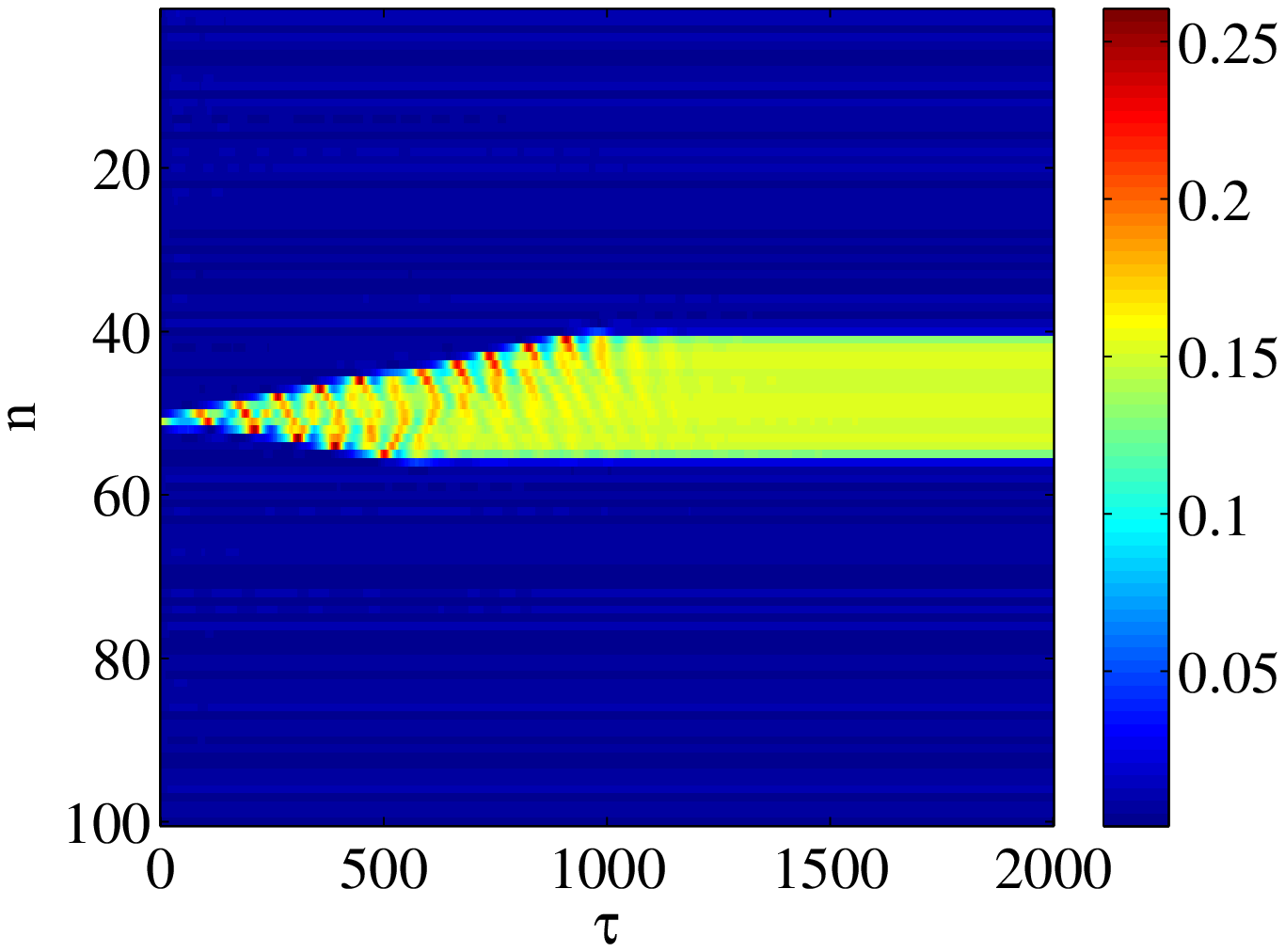}}\label{kappa4}
\subfigure[]{\includegraphics[width=4.3cm]{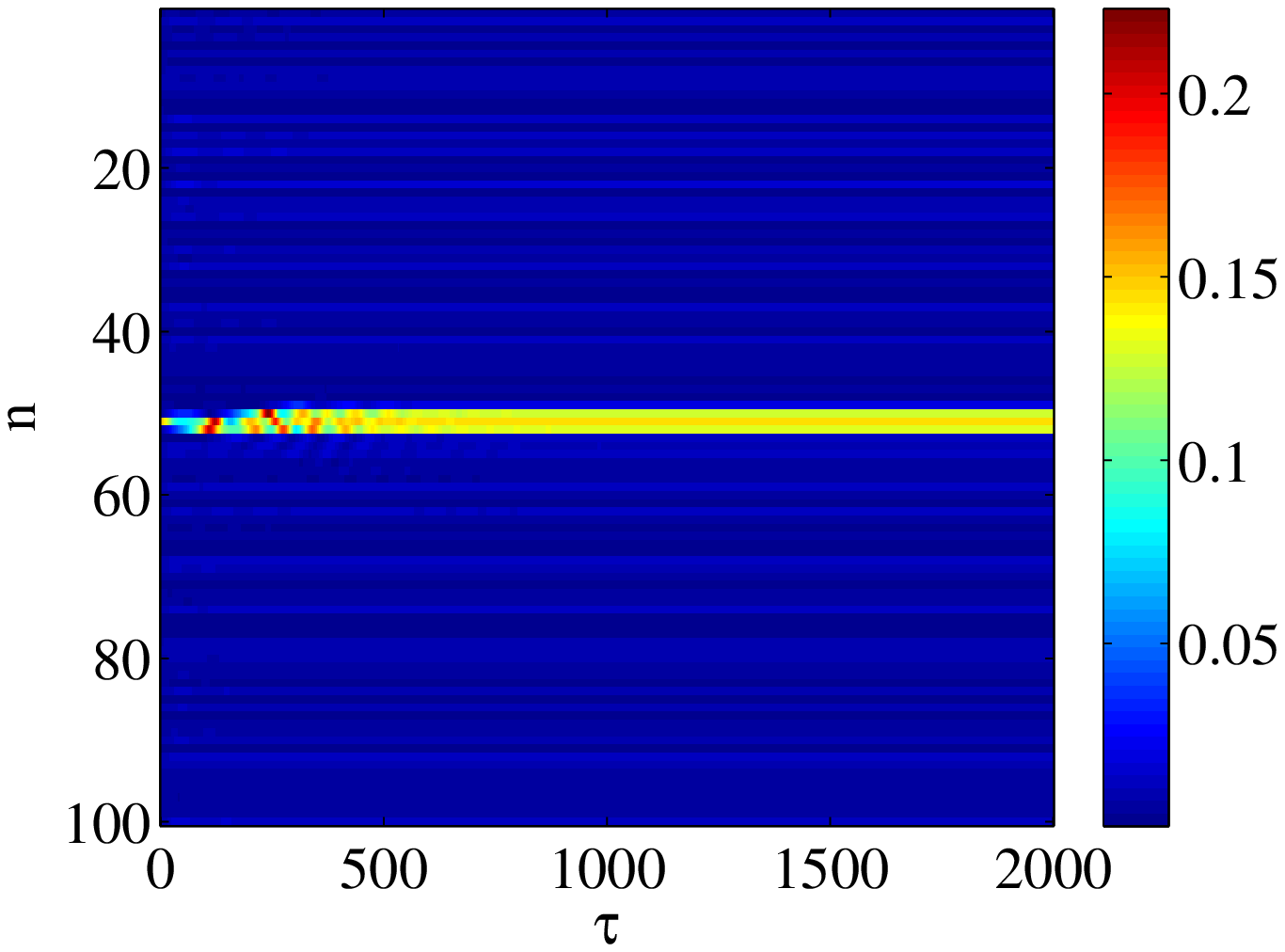}}\label{kappa6}
\subfigure[]{\includegraphics[width=4.3cm]{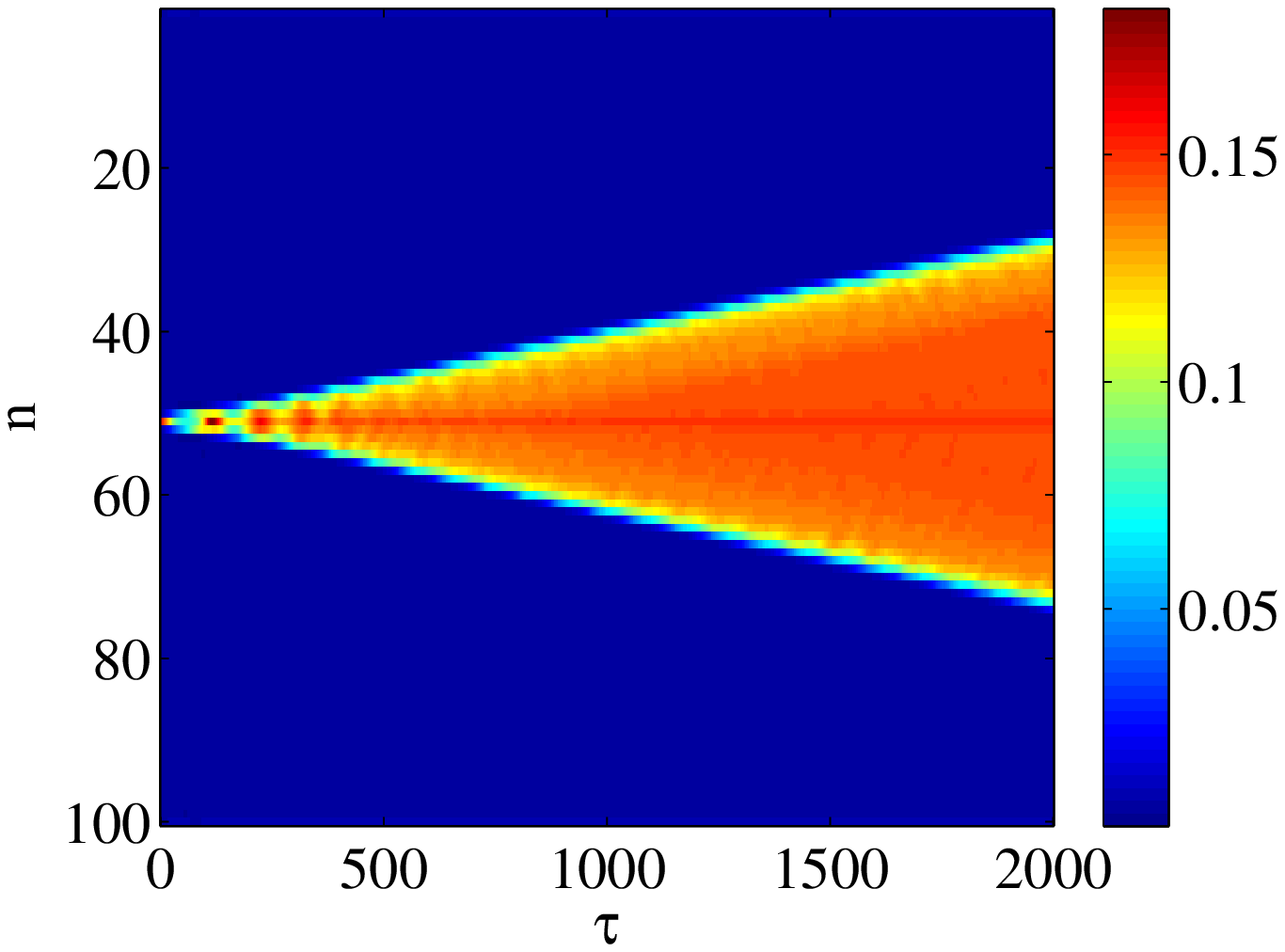}}\label{kappa2sta}
\subfigure[]{\includegraphics[width=4.3cm]{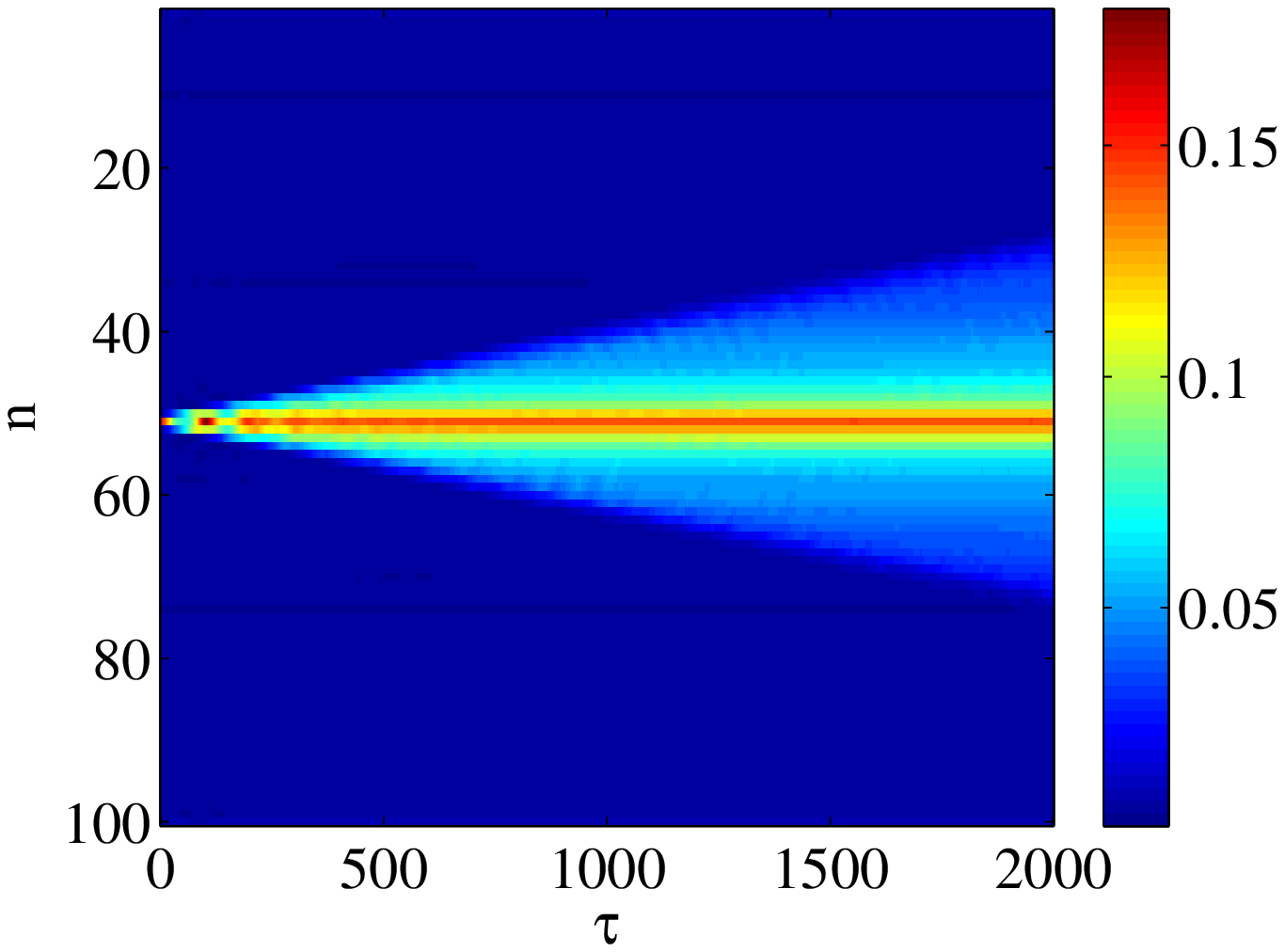}}\label{kappa4sta}
\subfigure[]{\includegraphics[width=4.3cm]{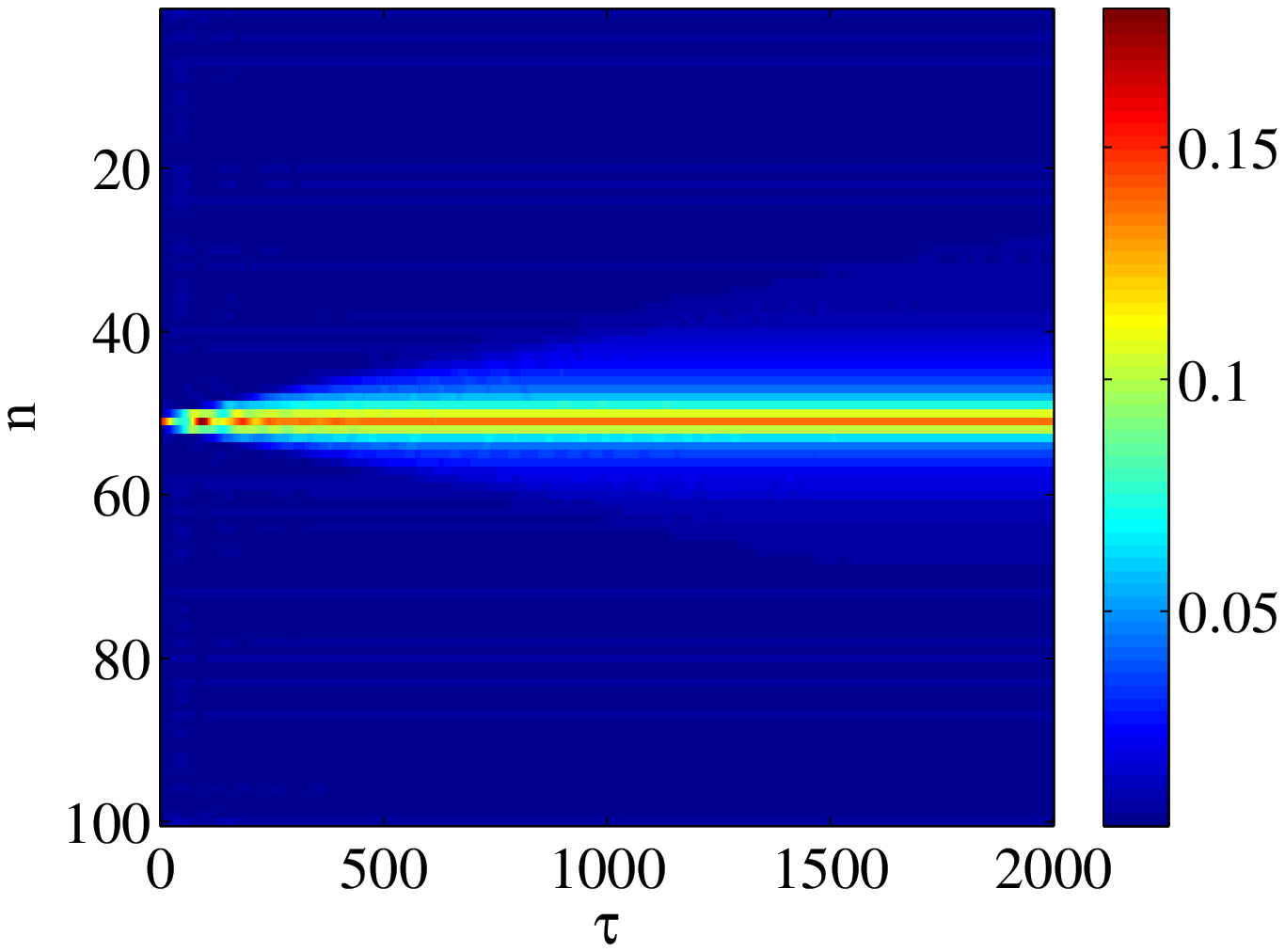}}\label{kappa6sta}
\caption{(a)$\sim$(c) The evolution of dipole intensity
  in a 100-particle system excited by the random electrical driving field under $\Omega=-0.1$, $|E_{0}|^2=0.9\times10^{-4}$: (a)
  $\kappa=0.2$, (b) $\kappa=0.4$, and (c)
  $\kappa=0.6$. (d)$\sim$(f) The corresponding statistical results for (a)$\sim$(c) after 200 real-time
  evolutions: (d) $\kappa=0.2$, (e) $\kappa=0.4$,
  and (f) $\kappa=0.6$.}
\end{figure}

\section{Extremely narrow Anderson localization}
We further study the effect of $\kappa$ on the size of the
Anderson-localized dipole mode. Here, we introduce the formula for
calculating the width of the dipole mode, which is written as
\begin{equation}
W=\frac{\left( \sum^{N}_{n=1}{\left| P_{n} \right|}^{2} \right)^{2}}{\sum^{N}_{n=1}{\left| P_{n} \right|}^{4}} \label{length},
\end{equation}
where $W$ denotes the width of the localized mode, and $N$ the number of nanoparticles. Note that in the finite
nanoparticle array, the value of the dipole that is located at the low
branch is not zero, which would give rise to calculation errors when
using Eq. (\ref{length}). To eliminate this error, ${P_{n}}$ should be
rescaled by reducing the average value of the dipole intensity that is located at the low branch.

Using equation (6), we can obtain the relation between $W$ and
$\kappa$, as shown in Fig. 6. The employed parameters are the same as
those shown in Fig. 5. Both Figs. 6(a) and 6(b) show that with a larger value of
$\kappa$, the value of $W$ becomes smaller, suggesting that the
statistical width of the localized dipole patterns becomes narrower. In particular, in the case of $\kappa=0.6$, the statistical width of the
dipole mode is approximately 7 particles (200 nm), which is in the scale of one incident wavelength, as shown from the violet curve in Fig. 6(b). The generated extremely narrow localized dipole mode in nanoparticle arrays may have
potential applications in highly precise detection and
manipulation of electromagnetic fields.

Further, it is interesting to find that there exists a threshold value $\kappa_{c}$, above which the AL of the dipole intensity can be generated, and below which the ballistic expansion of the dipole intensity occurs. The existence of $\kappa_c$ is resulted from the fact that there is a critical driving intensity $|E_{c}|^2$ for the kink moving. To generated localized patterns, the lower limiting value $|E_{0}|^2-\kappa|E_{0}|^2$ of the random driving intensity should be less than $|E_{c}|^2$. Based on this relation, we can find $\kappa_c=0.278$. Therefore, for the cases of $\kappa=0$ and $\kappa=0.2$, we cannot obtain the localized patterns, as shown in Figs. 5(a) and 5(d) and Fig. 6(b); while for the cases of $\kappa=0.4$ and $\kappa=0.6$, the localized patterns can be generated, as shown in Fig. 5 and Fig. 6.
\begin{figure}[htbp]
\centering
\subfigure[]{\includegraphics[width=6cm, height=5cm]{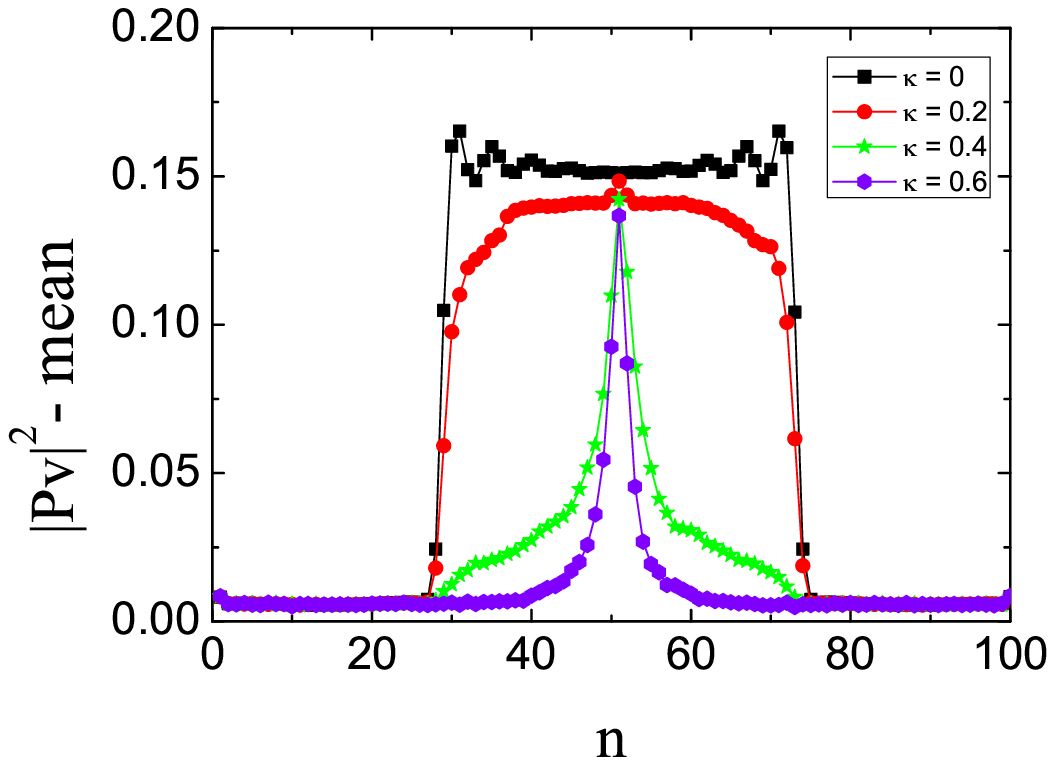}}\label{fig6anew}
\subfigure[]{\includegraphics[width=6cm, height=5cm]{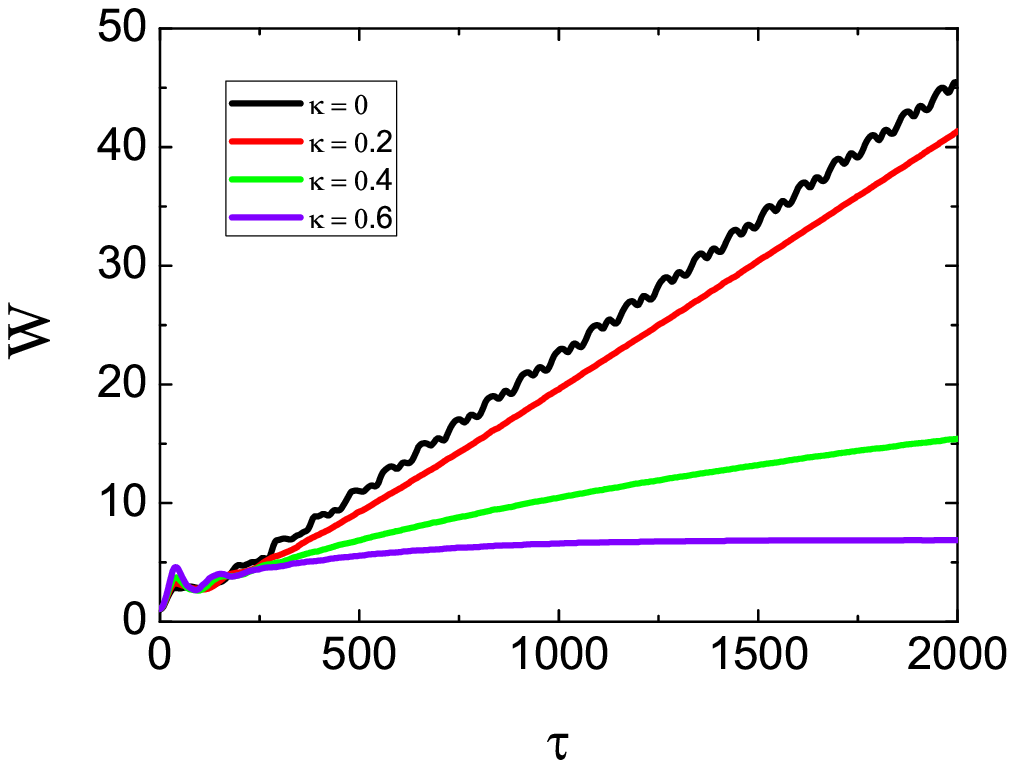}}\label{fig6bnew}
\caption{The statistical results of 200 real-time simulations with different values of $\kappa$. The detuning is set as $\Omega=-0.1$. (a) The distribution of mean value of the dipole intensity at $\tau=2000$; (b) the width evolution of localized dipole modes for different values of $\kappa$.}
\end{figure}

Obviously, $\kappa_{c}$ strongly relies on the external driving intensity $|E_{0}|^2$, as well as the critical driving intensity $|E_{c}|^2$ that is determined by the structural parameters, such as the radius of the metal sphere and the distance between two adjacent particles. For the present system we considered here (i.e., the critical driving intensity is fixed to $|E_{c}|^2=0.65\times10^{-4}$), the relation between $\kappa_c$ and $|E_{0}|^2$ can be determined by the expression $|E_{0}|^2-\kappa_c|E_{0}|^2=|E_{c}|^2$, with the result shown in Fig. 7. Figure 7 shows that increasing $|E_{0}|^2$ can also increases $\kappa_c$. This phenomenon can be understood as the following: the increase of the electrical driving intensity would strengthen the dipole-induced nonlinearity, and hence the expansion of the dipole intensity. As a result, to observe the localized modes, a stronger randomness (i.e., with a larger $\kappa$) of the electrical driving field is required.
\begin{figure}[htbp]
\centering
\subfigure[]{\includegraphics[width=7cm]{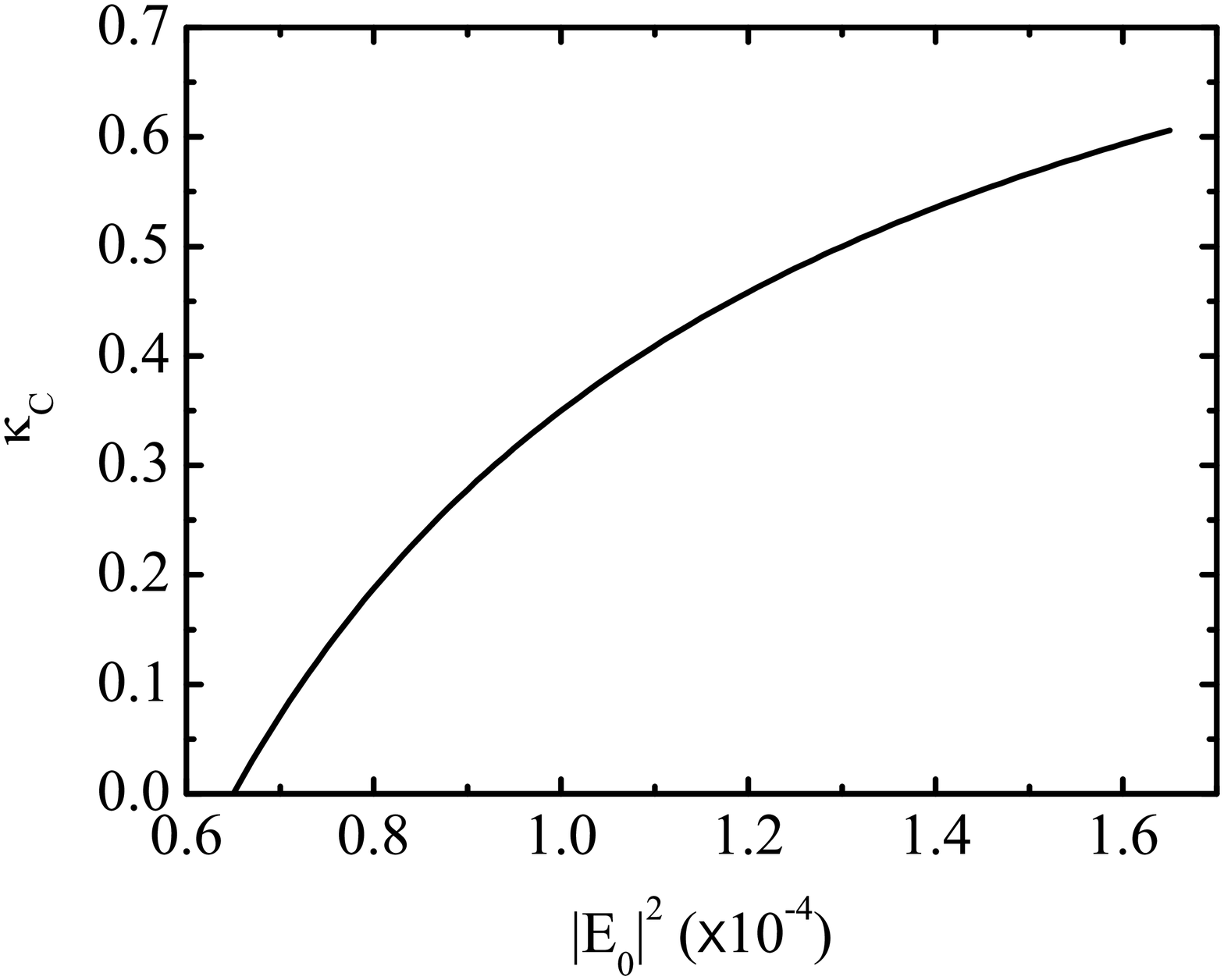}}\label{fig7}
\caption{The relation between $\kappa_{c}$ and the middle value of random external field($|E_{0}|^2$) with the same settings as discussed above.}
\end{figure}

\section{Conclusion}
In conclusion, we performed numerical analysis of Anderson localization of surface plasmon polaritons in a finite nonlinear
nanoparticle array. Unlike the conventional Anderson localization,
which is generally based on the structural disorder in a medium, here the random electrical driving field was utilized to excite the Anderson-localized patterns. It was shown that the dipole-induced nonlinearity results in ballistic expansion of dipole intensity; while the randomness of the external driving field can suppress such an expansion. Our numerical simulations by means of statistics confirmed that the effect of Anderson localization can be generated. We further demonstrated that the generated Anderson localization is highly confined, with its size down to the scale of one incident wavelength. Our results might provide a new scheme for the manipulations of electromagnetic fields in the scale of wavelength.

The possibility for experimentally generating the Anderson-localized dipole mode can be enhanced by reasonably adjusting the structure parameters such as increasing the radius of particle, as well as the center-to-center distance between two adjacent particles. Moreover, it is also possible to consider the random cell that contains two or more particles, which would significantly alleviate the requirements of driving field.

\begin{acknowledgments}
This work was supported, in part, by the National Natural Science Foundation of China through Grant Nos. 11575063, 61571197, 61172011 and 51505156.
\end{acknowledgments}
\appendix{Parameters}
The radius of the nanoparticle is fixed to  $a=10$ nm, and the distance between two adjacent particles is fixed to $d=30$ nm. An external optical field is
launched into the host and the dipole of the particles is excited. The dispersion of
the host can be neglected because the permittivity of SiO$_{2}$ is
nearly $\varepsilon_{h}\simeq2.15$ for the optical wavelength
range. The linear part of the dielectric constant of the silver
particle follows the Drude model, which can be expressed as
$\varepsilon_{\mathrm{Ag}}^{\mathrm{L}}=\varepsilon_{\infty}-\omega_{P}^{2}/[\omega(\omega-i\nu)]$,
where $\varepsilon_{\infty}=4.96$, $\hbar\omega_{P}=9.54$ eV and
$\hbar\nu$=0.055 eV \cite{oe18}. The nonlinear part of
the dielectric constant is selected as the standard cubic type, which
can be obtained as
$\varepsilon_{\mathrm{Ag}}^{\mathrm{NL}}=\chi^{(3)}|E_{n}|^{2}$, where
$\chi^{(3)}\simeq3\times10^{-9}$ esu for the silver spheres with a
radius of 10 nm\cite{oe20}, and $E_{n}$ is the local field inside the $n^{\mathrm{th}}$ particle. The frequency of the surface plasmon resonance of the silver nanoparticles, namely, $\omega_{0}$, can be expressed as $\omega_{0}=\omega_{P}/\sqrt{\varepsilon_{\infty}+2\varepsilon_{h}}$.

\bibliographystyle{plain}
\bibliography{apssamp}

\end{document}